\documentclass[reprint,prl,twocolumn,amsmath,amssymb]{revtex4-2}
\usepackage{graphicx}
\usepackage{dcolumn}
\usepackage{bm}
\usepackage{mathrsfs}  
\usepackage{braket} 
\usepackage{placeins}
\usepackage[svgnames]{xcolor}
\usepackage[colorlinks,citecolor=red,urlcolor=red,bookmarks=false,hypertexnames=true]{hyperref}
\usepackage{movie15}

\begin{document}

\preprint{APS/123-QED}

\title[Sample title]{Single-mode input squeezing and tripartite entanglement in three-mode ponderomotive optomechanics simulations}

\author{Kahlil Y. Dixon$^1$}
\email{kdixo16@lsu.edu}
\author{Lior Cohen$^2$}
\author{Narayan Bhusal$^1$}
\author{Jesse Frank$^1$}
\author{Jonathan P. Dowling$^1$}
\thanks{Deceased, June 5 2020.}
\author{Thomas Corbitt$^1$}
\affiliation{$^1$Department of Physics \& Astronomy, Louisiana State University, LA 70803, USA}
\affiliation{\mbox{$^2$Department of Electrical, Computer \& Energy Engineering, University of Colorado Boulder, CO 80309, USA}}


\date{\today}

\begin{abstract}
Quantum entanglement is a crucial resource for a wide variety of quantum technologies. However, the current state-of-art methods to generate quantum entanglement in optomechanical systems are not as efficient as all-optical methods utilizing nonlinear crystals. This article proposes a new scheme in which two single-mode squeezed light fields are injected into an optomechanical cavity. We demonstrate through our numerical simulations that the quantum entanglement can be substantially enhanced with the careful selection of squeezing strength and squeezing angle of the two quadrature squeezed light fields. Our results represent a significant improvement in output bipartite photon-photon entanglement over the previously demonstrated schemes using two coherent light fields as inputs. These simulations predict a maximum increase in bipartite optical entanglement by a factor of about 6, as well as increases in the quantum noise of the output light. A perceived loss of quantum information at certain squeezing angles is attributed to tripartite entanglement between the two optical fields and the optomechanical oscillator (OMO).  At particular squeezing angles, the bipartite (or tripartite) entanglement can be increased, thus introducing a method of optically controlling the intracavity entanglement. These mechanics can benefit various optical quantum technologies utilizing optomechanical entanglement and continuous variable quantum optics.

\end{abstract}

\maketitle

\section{\label{sec:level1}Introduction}
Quantum entanglement is one of the most important resources for a variety of quantum technologies such as quantum metrology \cite{giovannetti2011advances, dowling2002quantum}, quantum communication \cite{khatri2020principles}, quantum sensing \cite{PhysRevA.78.063828}, quantum simulation \cite{QLGSIM}, and quantum imaging \cite{Maga_a_Loaiza_2019}. Producing entanglement via exploitation of the three-wave mixing process in optomechanical cavities (OMC) is an attractive and innovative process for generating entangled photons \cite{Wipf_2008,PhysRevResearch.2.033244}. However, the current performances of optomechanical cavities to produce entanglement are not as robust as other methods utilizing nonlinear crystals \cite{Chen2020}. 

Our previous study has suggested that ponderomotive (optomechanical) entanglement is highly dependent on input quantum noise \cite{Kahlil2020}. Conveniently, squeezed light\textemdash a non-classical state of light widely used in quantum optical applications\textemdash has suppressed/enhanced quadrature noise that can be tailored based on the squeezing strength and squeezing angle \cite{teich1989squeezed, matekole2020quantum, gerry_knight_2004}. In order to further investigate the relationship between quantum noise and the strength of pondermotive entanglement, we inject such a light field, squeezed coherent light. The primary motivation of this project is to study the effect of the input quantum noise on the nonlinear dynamics in the optomechanical cavity (OMC).  We expected that, under certain conditions, the output bipartite optical entanglement should also be enhanced by the input squeezing \cite{dixon2020optomechanical}. 
Furthermore, there are now several publications discussing non-Gaussian output from OMCs, with only Gaussian inputs  \cite{PhysRevA.56.4175,Aspelmeyer2014,Qvarfort_2019,Qvarfort_2020}. This phenomenon is only reported in systems that exhibit the full $\chi^{(3)}$ nonlinear interaction or in linearized systems that exhibit strong quantum coupling \cite{Aspelmeyer2014,PhysRevA.56.4175}. This is not the case for these simulations. Thus, we utilize these Gaussianity metrics to examine the loss of information when measuring two of three modes in a system with entanglement the fluctuates between bipartite and tripartite entanglement \cite{PhysRevLett.125.020502,PhysRevX.10.011011}.

We report (``linearized") simulations of an optomechanical cavity in the unresolved sideband regime that exhibits intracavity quantum noise dominance (though in the weak quantum coupling regime) \cite{dixon2020optomechanical,aggarwal2018room,Leijssen2017}. 
Moreover, the injection of single-mode squeezed light  enhances both optical bipartite entanglement output from optomechanical cavity and tripartite entanglement of the intracavity quantum states \cite{PhysRevResearch.2.033244}. These simulations assume sideband squeezing of the input coherent light into an optomechanical cavity  with a small movable end mirror optomechanical oscillator ( rest mass of $\approx 50$ng). These simulations are performed in a parameter regime that yields bipartite optical ponderomotive entanglement without input squeezing (the intracavity quantum noise dominance regime) (squeezing strength $r$ equal to zero) \cite{sharifi2019design,Kahlil2020,dixon_2021}.  

 After considering limited resources, these devices may be competitive to conventional entanglement methods in certain conditions, especially those that can exploit entanglement in continuous-variable quantum states.

\begin{figure}[!ht]
    \centering
    \includegraphics[width=0.48\textwidth]{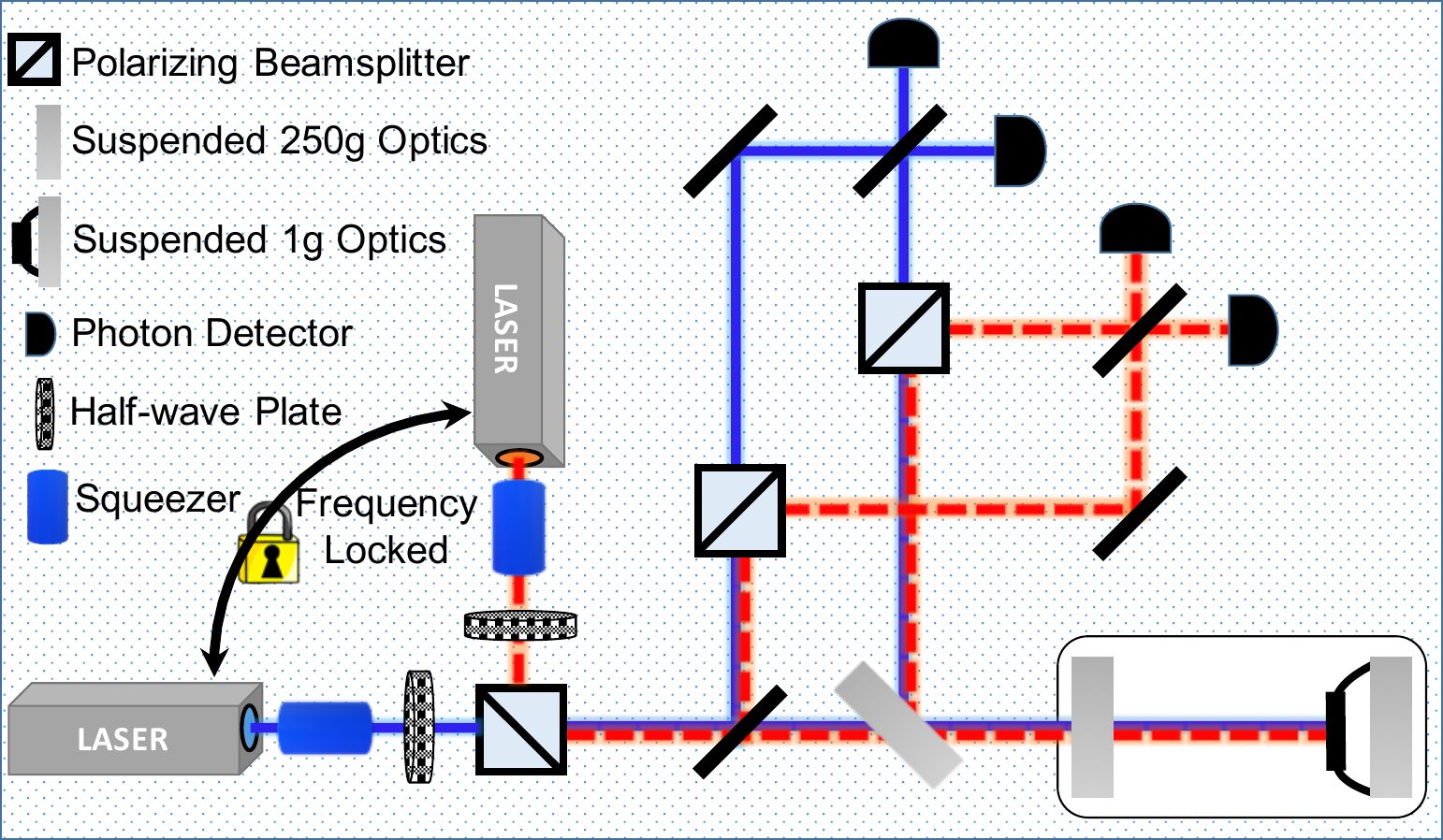}
    \caption{Schematic diagram of the proposed experimental setup. The two frequency-locked pump lasers and nonlinear crystals produce two single-mode squeezed light fields. The two quadrature squeezed light fields are injected into the optomechanical cavity. Balanced-homodyne measurements are proposed to quantify the output fields' entanglement from the optomechanical cavity. The two input fields are detuned from the same cavity resonance (carrier detuning coefficient 0.3 and -1.5 for the subcarrier).}
    \label{fig:setup}
\end{figure}


\section{\label{sec:level1}Ponderomotive quantum coupling}

Optomechanics is critical to a plethora of precision devices and instruments \cite{Kasunic_2015}. Optomechanics (and Cavity Optomechanics) applications are a go-to solution for classical optics-based technologies and an emerging technology for Photonics and continuum Quantum Optics applications \cite{9430488,Carney_2021}. Optomechanical devices are optical devices that contain a mechanical oscillator coupled to the propagating optical modes. Optomechanics even appears as an attractive method for cutting edge quantum computing devices such as quantum memory and quantum teleportation for quantum computing IO device development \cite{neuman2020phononic}. The utility of optomechanics will further expand into Quantum Information Science technologies as a means of producing quantum states and quantum measurements in various applications.

The simulations all consider the optical apparatus depicted in figure \ref{fig:setup}. Two coherent states are squeezed by arbitrary black-box squeezers that impart the squeezing parameters, $\xi_1$ and $\xi_2$ (where $\xi_j=r_j e^{i\theta_j}$), on their respective coherent state. For the purpose of this report, consider these squeezers to have easily tunable squeezing strength ($r_1$ and $r_2$) and squeezing angle ($\theta_1$ and $\theta_2$); at times swap out the squeezing strength $r_j$ to a tunable nonlinearity coefficient, $\mu_1$ and $\mu_2$ (where $r_j=\sqrt{P_j}\mu_j$), to better consider management (maximization) of quantum resources. The optomechanical cavity is a cantilever shaped movable $\mu$mirror \cite{cripe_2018,sharifi2019design}. Which has higher-order vibrational modes that affect output entanglement \cite{Kahlil2020,dixon_2021}. Though there is some discussion of sideband frequency dependence in the appendix, the sideband frequency dependence in these simulations follows the same behavior as the unsqueezed case discussed in a previous publication. 

First, consider two unsqueezed coherent states input into the cavity that may yield bipartite entangled optical mixed quantum states. The Hamiltonian for dual coherent fields incident onto an optomechanical cavity with a $\mu$mirror oscillator (movable endmirror with small mass) has the following optical-mechanical interaction term:
\begin{eqnarray}\
H_{int}= \hbar (g_1\hat{\mathbf{n}}_1+g_2\hat{\mathbf{n}}_2)\sum_k(-1)^k\frac{(\hat{\mathbf{X}}_\mu/L-1)^k}{k!}~,
\label{eqn:H_int}
\end{eqnarray}
where $g$ is some optomechanical coupling constant, $\hat{\mathbf{X}}_\mu$ is the position quadrature operator corresponding to the $\mu$mirror, $\hat{\mathbf{n}}_j= \hat{\mathbf{a}}^\dagger_j\hat{\mathbf{a}}_j$ is the photon number of the $j^{th}$ intracavity field, and L is the cavity length when input quantum noise is zero~\cite{wipf_2013}.

\subsubsection{Linearized Hamiltonian}
The linearization provides a more intuitive Hamiltonian and is valid in the simulation's regime. The goal is to write the interaction Hamiltonian using the fluctuations in the optical cavity field around a steady-state average coherent amplitude, $\alpha_j$. For a cavity driven by coherent fields $\ket{\alpha_j}$ this involves the substitution $\hat{\mathbf{a}}_j\rightarrow \alpha_j+\delta\hat{\mathbf{a}}_j$ into the Hamiltonian.
\begin{equation}
\begin{aligned}
H_{int} \approx (g'_1\hat{\mathbf{n}}_1+g'_2\hat{\mathbf{n}}_2) \hat{\mathbf{X}}_\mu~\\
H_{int}^{lin} =  \sum_j g'_j (\delta\hat{\mathbf{a}}^\dagger_j+\alpha_j^*)(\delta\hat{\mathbf{a}}_j+\alpha_j) \hat{\mathbf{X}}_\mu~\\
H_{int}^{lin} = \sum_j g'_j (\delta\hat{\mathbf{a}}^\dagger_j\delta\hat{\mathbf{a}}_j+|\alpha_j|^2+\alpha_j\delta\hat{\mathbf{a}}^\dagger_j+\alpha^*\delta\hat{\mathbf{a}}_j) \hat{\mathbf{X}}_\mu~
\end{aligned}
\end{equation}
Take the new rest position of the mechanical oscillator to be displaced by the average radiation pressure force, $g'_j|\alpha_j|^2\langle{\hat{\mathbf{X}}_\mu}\rangle~$. Then,
\begin{equation}
    H_{int}^{lin} = \sum_j g'_j (\delta\hat{\mathbf{a}}^\dagger_j\delta\hat{\mathbf{a}}_j+\alpha_j\delta\hat{\mathbf{a}}^\dagger_j+\alpha_j^*\delta\hat{\mathbf{a}}_j) \hat{\mathbf{X}}_\mu~.
\end{equation}
In cases of weak quantum coupling, the term $g'_j\delta\hat{\mathbf{a}}^\dagger_j\delta\hat{\mathbf{a}}_j$ is negligible since the other terms have factors  $\alpha_j$. Thus the Hamiltonian simplifies to,
\begin{equation}
\begin{aligned}
     H_{int}^{lin} =  \sum_j g'_j (\alpha_j\delta\hat{\mathbf{a}}^\dagger_j+\alpha_j^*\delta\hat{\mathbf{a}}_j) \hat{\mathbf{X}}_\mu~\\
     H_{int}^{lin} =  \sum_j g'_j (\alpha_j\delta\hat{\mathbf{a}}^\dagger_j+\alpha_j^*\delta\hat{\mathbf{a}}_j) (\hat{\mathbf{b}}^\dagger + \hat{\mathbf{b}})~,
\end{aligned}
\end{equation}
where $x_{zpf}g'_j\rightarrow g'_j$.
In our simulations the two input fields are blue and red-detuned from the cavity resonance. This means that the dominate terms in the above Hamiltonian become the following:
\begin{equation}
    g'_1 \alpha_1\delta\hat{\mathbf{a}}^\dagger_1\hat{\mathbf{b}}^\dagger + g'_2 \alpha_2\delta\hat{\mathbf{a}}^\dagger_2 \hat{\mathbf{b}} +  g'_1 \alpha_1^*\delta\hat{\mathbf{a}}_1\hat{\mathbf{b}} + g'_2 \alpha_2^*\delta\hat{\mathbf{a}}_2 \hat{\mathbf{b}}^\dagger~.
\end{equation}
These competing photon-phonon processes (for example the input relation for the carrier field, $g'_1 \alpha_1\delta\hat{\mathbf{a}}^\dagger_1\hat{\mathbf{b}}^\dagger $, \textit{competes} with the input relation for the subcarrier field $g'_2 \alpha_2\delta\hat{\mathbf{a}}^\dagger_2 \hat{\mathbf{b}}$ and likewise for the outputs) will cause \textit{additional} fluctuations in the mean cavity coherent amplitudes, $\alpha_j$; though not in this linearized simulation, previous work has shown that a single cavity field has an intricate dependence on the change in cavity linewidth \cite{PhysRevA.73.023801}. However, in this simulation, the cavity photon fluctuations can be seen in the sideband photon numbers ($\delta\hat{\mathbf{a}}_j^\dagger \delta\hat{\mathbf{a}}_j$), though not significant enough to move the system into the quantum coupling regime.

The optomechanical coupling modulation could also be due to two-tone optomechanical instabilities that occur within particular choices of carrier and subcarrier field detunings in the unresolved sideband regime \cite{PhysRevA.103.023525, JSZhang20,PhysRevX.9.041022}.

\section{Squeezed input boosts entanglement}

Following the Quantum Langevin method, the Langevin equation for squeezed light injection is very similar to the unsqueezed OM entanglement calculations in previous publications \cite{AuBougouffa2020,Ghasemi_2021}. Since the coupling matrix, $\mathbf{K}$, does not change for lossless squeezing, we only consider alterations to the input noise spectra $\langle \mathbf{G} \rangle$ \cite{Asjad:19,Qvarfort_2020}. Recall that the quantum Langevin equations calculate  (in Heisenberg's picture of quantum mechanics) the time derivative of the intracavity quadrature operators. Let $\mathbf{u}_c$ be the vector of intracavity mode operators, and the vector $\mathbf{u}_N$ be the vector of input mode operators. Then, the intracavity modes change in time according to the relation:
\begin{eqnarray}
\dot{\mathbf{u}}_c= \mathbf{K}\mathbf{u}_c+ \mathbf{u}_N~.
\end{eqnarray} After Fourier transforming to the sideband frequency $\Omega$, applying cavity input-output theory to establish the vector of quadratures of the output optical fields $v$, and arranging the terms into a covariance matrix, we see that the correlations in the input noises $\langle \mathbf{G} \rangle$ evolves into the output optical correlations, covariance matrix $\mathbf{V}$:
\begin{equation}
\mathbf{V}(\Omega) = Re(\mathbf{Q}\langle{\mathbf{G}}\rangle \mathbf{Q}^\dagger),
\end{equation}
where $\mathbf{Q}= \sqrt{2\gamma_c}\mathbf{M}(\Omega)+\mathbf{I}$, and $\mathbf{M}(\Omega)= [\mathbf{K}+i\Omega \mathbf{I}]^{-1}$ .

To quantify the entanglement with the logarithmic negativity $E_N$ begin with the direct computation of the elements of the quadrature covariance matrix, $\mathbf{V}$: 
\begin{equation}
E_N= \max[~0, -\ln{\sqrt{2\eta -2\sqrt{\eta^2-4 \det\mathbf{V}}}}],
\end{equation}
where $\eta=\det V_{11}+ \det V_{22} -2\det V_{12}$. This is a popular metric for Gaussian entanglement in continuous variable systems. One reason for its popularity is that for quantum Gaussian states, logarithmic negativity amounts to an exact entanglement cost \cite{PhysRevLett.125.040502,wang2018exact}.

Note that the quadrature covariance matrix is of the form: 
\begin{eqnarray}
\hspace*{-0.5cm}\mathbf{V}=\left(
\begin{array}{cccc}
\left\langle  {\mathbf{X}}_1   {\mathbf{X}}_1\right\rangle_+  & \left\langle
    {\mathbf{X}}_1   {\mathbf{Y}}_1\right\rangle_+  & \left\langle
    {\mathbf{X}}_1   {\mathbf{X}}_2\right\rangle_+  & \left\langle
    {\mathbf{X}}_1   {\mathbf{Y}}_2\right\rangle_+  \\
 \left\langle  {\mathbf{Y}}_1   {\mathbf{X}}_1\right\rangle_+  & \left\langle
    {\mathbf{Y}}_1   {\mathbf{Y}}_1\right\rangle_+  & \left\langle
    {\mathbf{Y}}_1   {\mathbf{X}}_2\right\rangle_+  & \left\langle
    {\mathbf{Y}}_1   {\mathbf{Y}}_2\right\rangle_+  \\
 \left\langle  {\mathbf{X}}_2   {\mathbf{X}}_1\right\rangle_+  & \left\langle
    {\mathbf{X}}_2   {\mathbf{Y}}_1\right\rangle_+  & \left\langle
    {\mathbf{X}}_2   {\mathbf{X}}_2\right\rangle_+  & \left\langle
    {\mathbf{X}}_2   {\mathbf{Y}}_2\right\rangle_+  \\
 \left\langle  {\mathbf{Y}}_2   {\mathbf{X}}_1\right\rangle_+  & \left\langle
    {\mathbf{Y}}_2   {\mathbf{Y}}_1\right\rangle_+  & \left\langle
    {\mathbf{Y}}_2   {\mathbf{X}}_2\right\rangle_+  & \left\langle
    {\mathbf{Y}}_2   {\mathbf{Y}}_2\right\rangle_+  \\
\end{array}
\right)
\end{eqnarray} 
were $\left\langle u^*v\right\rangle_+~=~\frac{\left\langle u^*v+ v^*u\right\rangle}{2}$, 
or in block form:
\begin{eqnarray}
\centering
\mathbf{V}= \left(
\begin{array}{cc}
 V_{1 1} & V_{1 2} \\
 V_{2 1} & V_{2 2} \\
\end{array}
\right).
\end{eqnarray}
Of the 16 terms in the optical output covariance matrix, the four diagonal terms are (somewhat) trivial, and six others can be calculated directly from the remaining six. Therefore, we only need to calculate six terms to construct the matrix (see appendix). The six terms can be written and calculated a six products terms of the form \cite{adhikari2018phase}:
\begin{eqnarray}
\hat{\mathbf{S}}_\ell^\dagger( \xi_\ell) \hat{\mathbf{Y}}_\ell\hat{\mathbf{S}}_\ell( \zeta_\ell) = \hat{\mathbf{S}}_\ell^\dagger(\xi_\ell-\zeta_\ell) \hat{\mathbf{\gamma}}_y( \zeta_\ell)~,
\end{eqnarray} with $\zeta = r'e^{i\theta'}$, and $\hat{\mathbf{\gamma}}_y( \zeta_\ell) = \frac{i}{2}[(\cosh{ r'_\ell}+e^{i \theta'_\ell}\sinh{ r'_\ell})\hat{\mathbf{a}}^\dagger-(\cosh{r'_\ell}+e^{-i \theta'_\ell}\sinh{r'_\ell})\hat{\mathbf{a}}]$. 

\begin{equation}
\begin{aligned}
&\langle{\hat{\mathbf{G}}}\rangle_{j,j}=\\
&\resizebox{0.42\textwidth}{!}{$\left(
\begin{array}{cc}
 \frac{A_j^2\alpha+B_j^2\alpha^{*2}+A_jB_j(2|\alpha|^2+1)}{2} & \frac{i(B_j^2 \alpha^{*2}-A_j^2\alpha^2 + A_jB_j)-1}{2}\\
 -\frac{i(B_j^2 \alpha^{*2}-A_j^2\alpha^2 + A_jB_j)-1}{2} & \frac{-A_j^2\alpha-B_j^2\alpha^{*2}+A_jB_j(2|\alpha|^2+1)-1}{2} 
\end{array}\right)$}
\end{aligned}
\end{equation} where $A_j= \cosh{r_j}-e^{-i\theta_j}\sinh{r_j}$ and $B_j= \cosh{r_j}-e^{i\theta_j}\sinh{r_j}$.

\begin{figure}
    \centering
    \includegraphics[width=8.60cm]{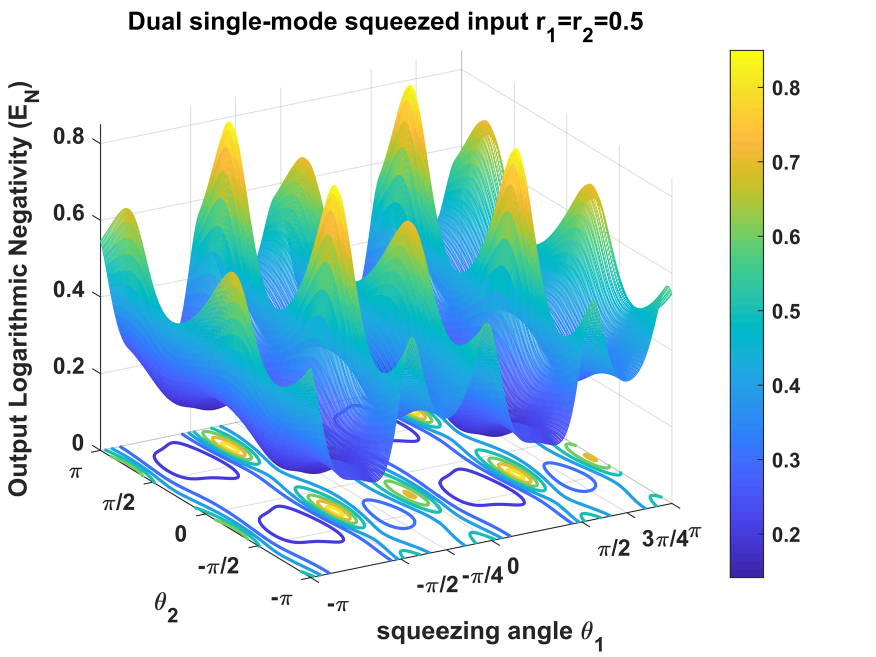}
     \includegraphics[width=8.60cm]{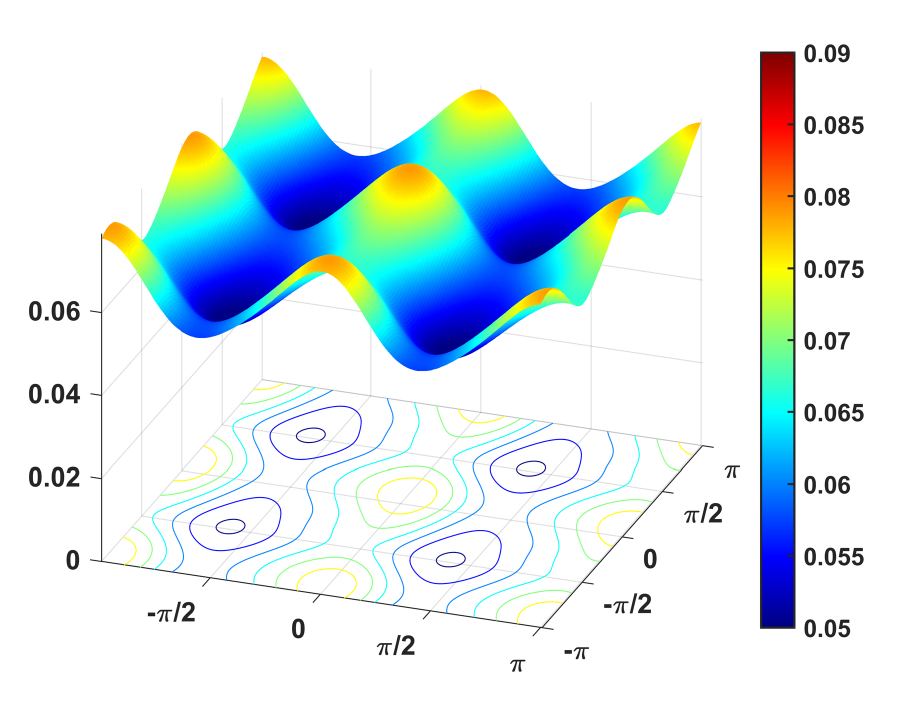}
    \caption{The simulation shows the same two-mode periodicity due to the dependence on two squeezing angles. This dependence is consistent with our presented theoretical results Quantum Langevin calculations. The maximum output $E_N$ varies strongly with input squeezing and angle. Temperature is set to $4K$ (the simulation fails at room temperature), and see table \ref{t1} for other parameters used. The maxima occur when input squeezing angles are odd integer multiples of $\pi/2$  and minima at even integer multiples of $\pi/2$. The last plot in the figure shows the results from applying Duan's measure in the same simulated conditions. Their discrepancy is due to non-Gaussianity in the output fields.}
    \label{21msthetas}
\end{figure}


The two measures of entanglement used here are only ideal for continuous-variable bipartite states. Notably, the two measures (logarithmic negativity and Duan's measure of inseparability) do not agree (figure \ref{21msthetas}). The partial measurement of the states produced by these simulations obscures one of three entangled modes; the input squeezing of the optical fields then modulates that obstructed tripartite entanglement. Consideration of realistic conditions (such as losses, thermal noise, more realistic OMO dampening, and higher-order mechanical modes in the OMO) are included in most of the simulations (see figure \ref{21msthetas}). 

\section{Examining the entanglement with Gaussianity metrics}

\begin{figure}
    \centering
    \includegraphics[width=4.3cm]{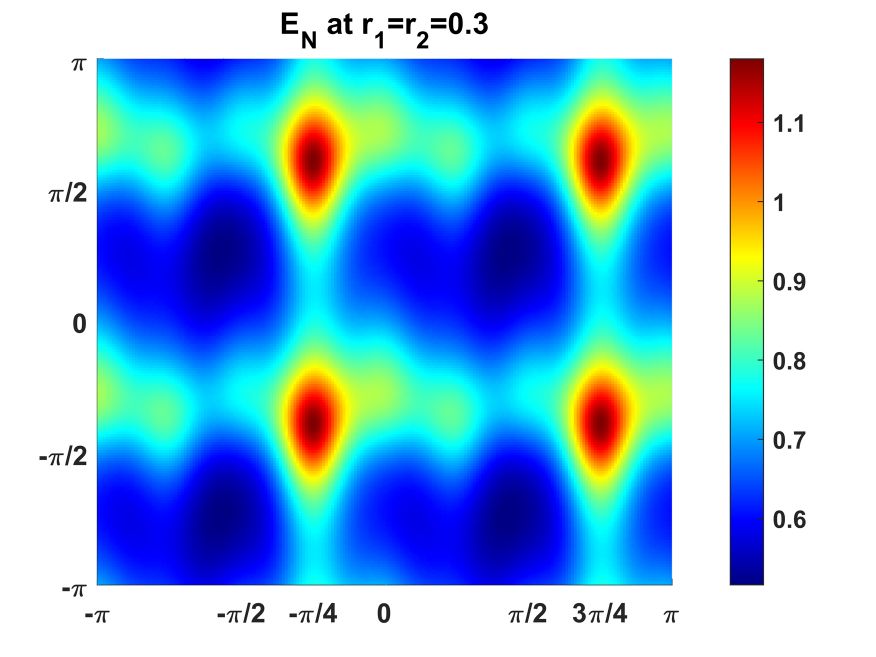}\includegraphics[width=4.3cm]{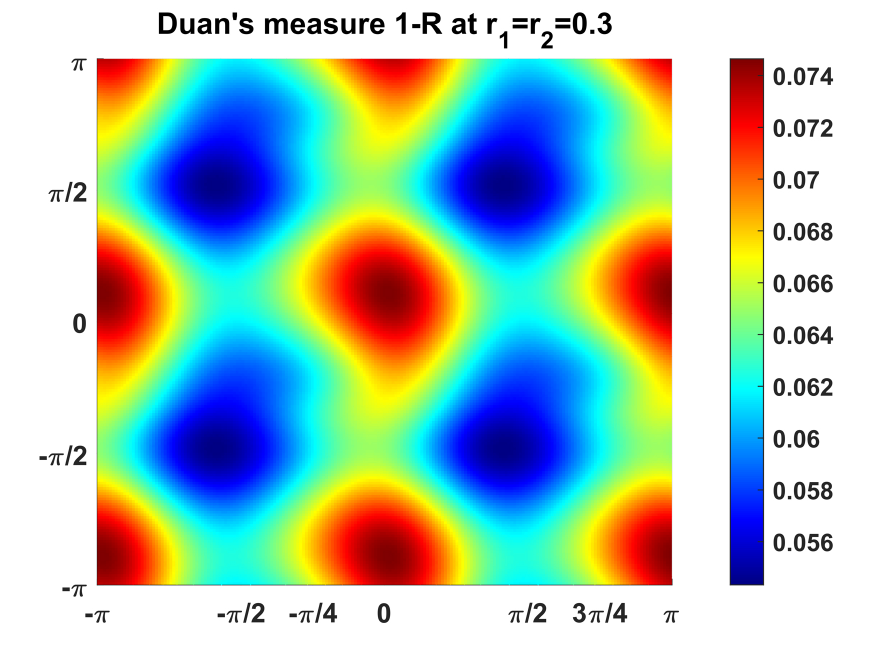}\\
        \includegraphics[width=4.3cm]{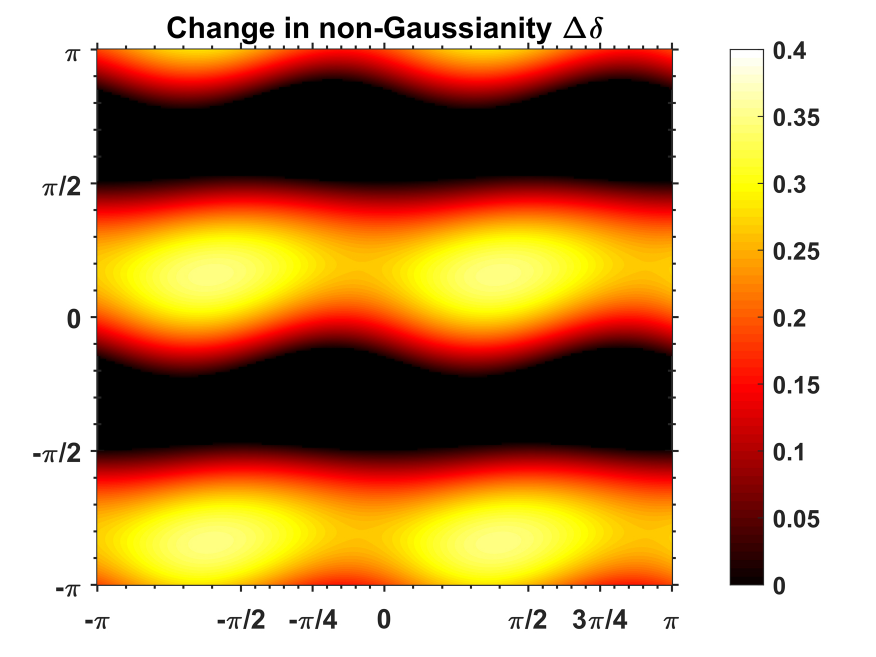}
    \caption{All three plots are at zero loses and $10$nK and equal squeezing strengths ($r_1=r_2=0.3$) to simplify the measures of Gaussianity and show changes with squeezing angle. A: The top left plot is the logarithmic negativity ($E_N$). The logarithmic negativity corresponds to the bipartite entanglement between the optical fields. B: The top right plot depicts results from Duan's measure of separability $R$ rewritten here as $1-R$ to be positive for entanglement and zero or less for any separable state. Duan's measure of separability is more reliable than the logarithmic negativity when involving continuous variable entanglement beyond bipartite entanglement.
    The peaks in this plot occur at different angles versus the $E_N$. If entanglement with the OMO is considered better by Duan's measure then the changes Duan's measure will coincide with changes in the Gaussianity metric. This is the case. C: The lower plot is the measured change the Gaussianity measure ($\delta$) in the output optical fields only. Here, this measure does not indicate the Gaussianity but the loss of information due to entanglement of one or both optical fields with the mechanical oscillator. The higher the value, the more information is hidden to the partial measurement of the system. This implementation of this metric only holds for pure states, so we have set temperature and loss to their minimums for this figure. The angles for lower $\Delta\delta$ directly coincide with the regions of maximum logarithmic negativity. This suggests that when these systems are tuned away from tripartite entanglement towards bipartite entanglement between the propagating optical fields only, the output bipartite entanglement is boosted. }
    \label{fig:Gaussianity}
\end{figure}

To further study the non-Gaussianity of the output optical fields we use the Genoni's metric for deviation from Gaussianity for ideal systems \cite{PhysRevA.78.060303,PhysRevA.76.042327,PhysRevA.82.052341}:

\begin{equation}
    \delta(\hat{\rho})= S(\rho_G) \geq S(\mathbf{V})-S(\mathbf{G})-S(\hat{\rho}_{\mu 0})
\end{equation}
where $S(\mathbf{V})$ is the Von Neumann entropy of $\mathbf{V}$, the covariance matrix of $\rho_G$ is $\mathbf{V}$ (and thus the covariance matrix of the output optical fields) in standardized form \cite{PhysRevLett.84.2722}, $\mathbf{G}$ is the covariance matrix of the input optical fields mentioned previously, and $\rho_G$ is the Gaussian description of the state $\rho$. The Von Neumann entropy of a Gaussian system is a function of the covariance matrix's symplectic eigenvalues \cite{wilde_2013}. Figure \ref{fig:Gaussianity} shows the difference in Guassianity,
\begin{equation}
    \Delta \delta=\delta(\hat{\rho}(\xi_1,\xi_2)) - \delta(\hat{\rho}(0,0))~.
\end{equation}

In Figure \ref{fig:Gaussianity}, entanglement measure results are also included.  All three plots are at zero losses and $10$nK and equal squeezing strengths ($r_1=r_2=0.3$) to simplify Gaussianity measurement and to show dependence on squeezing angle. Nonzero logarithmic negativity is indicative of bipartite entanglement between the propagating fields; this entanglement can be controlled via adjustments to the squeezing strength and angles. Duan's measure is minimum where the Logarithmic negativity is maximum, despite both being measures of entanglement. Since the $E_N$ does not consider the higher-order statistics, the maxima in Duan's measure must be due to the changes in the entanglement entropy that persists in system at certain optical squeezing parameters. The results from applying Genoni's Gaussianity metric reveal these parameters. Genoni's measure will be positive not only when there is tripartite entanglement but when there is any entanglement between a measured optical field and the OMO. This is because it measures how poorly the pure three-mode state was approximated by the two-mode Gaussian state created from the measurement data. Specifically, here it measures how much information is lost or gained ( to or from the OMO ) when squeezing is applied to the input fields versus without squeezing. Consequentially, minima in this measure highlight where the squeezing reduces entanglement with the OMO, these are regions where the squeezing can boost the $E_N$ relative to unsqueezed input. These regions must correspond to weaker tripartite entanglement as well. These results are consistent with the quantum noise results, when input quantum noise is maximum entanglement (including tripartite entanglement) is maximum.

\FloatBarrier

\section{OMC entanglement device efficacy}


\begin{table}[!ht]
\centering
$\begin{array}{|c|c|c|}
 \text{Parameter} & \text{Notation} & \text{Stable and $E_N \neq 0$} \\
 \hline
 \text{Temperature} & T & 4 \text{K} \\
 \text{Circulating carrier } \text{power} & P_1 &
   0.2816 \text{W} \\
 \text{Circulating subcarrier power} &
   P_2 & 0.2238 \text{W} \\
 \text{Loss} & L_s & 25ppm \\
 \text{Carrier } \text{detuning} &
   d_1 & 0.3 \\
 \text{Subcarrier detuning} &
   d_2 & -1.5 \\
 \text{Quality factor} & Q &
   17000 \\
 \text{Cavity } \text{Length} & L_n &
   0.01 \text{m} \\
\end{array}$
\caption{Set of variable simulation parameters, and a stable configuration that yield non-zero entanglement.  }
\label{t1}
\end{table}

\subsection{Quantum Noise}

Two coherent states input into an optomechanical cavity has been shown theoretically and experimentally to generate entanglement \cite{Chen2020, Wipf_2008, PhysRevLett.98.030405, dixon2020optomechanical}.  Moreover, recent work has provided more insight into what drives this quantum process. Specifically, the relationships between output entanglement and input quantum noise; not only does the output entanglement thrive when the quantum noise is dominant in the optomechanical cavity, but that, at low thermal noise (lower than room temperature for this system), the quantum noise itself is more indicative of the output entanglement than the ratio. Thus, to increase the entanglement output from the optomechanical cavity, we need to increase the input quantum noise while keeping the mechanical oscillator's temperature reasonably low. For the oscillator to maintain the low relative classical noise, the input light must remain at or near minimum uncertainty at the input; increasing the fluctuations in the input light's amplitude consequentially decreases the fluctuations in that light's phase to maintain this quantum only input noise~\cite{gerry_knight_2004}.  
Thus, we intentionally amplify the intracavity quantum noise by injecting displaced squeezed states (or squeezed coherent light).

\begin{figure}
    \centering
    \includegraphics[width=4cm]{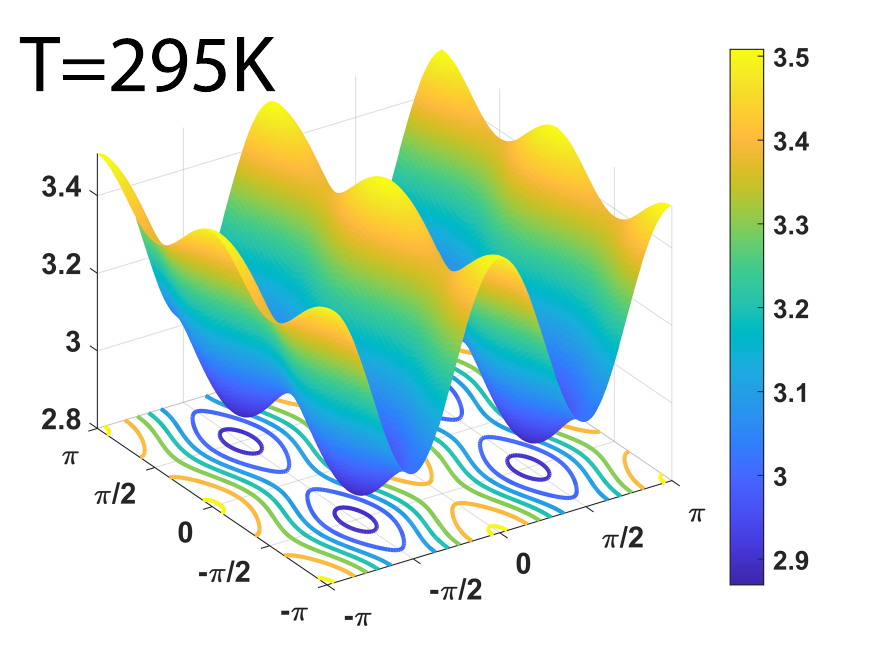}\includegraphics[width=4cm]{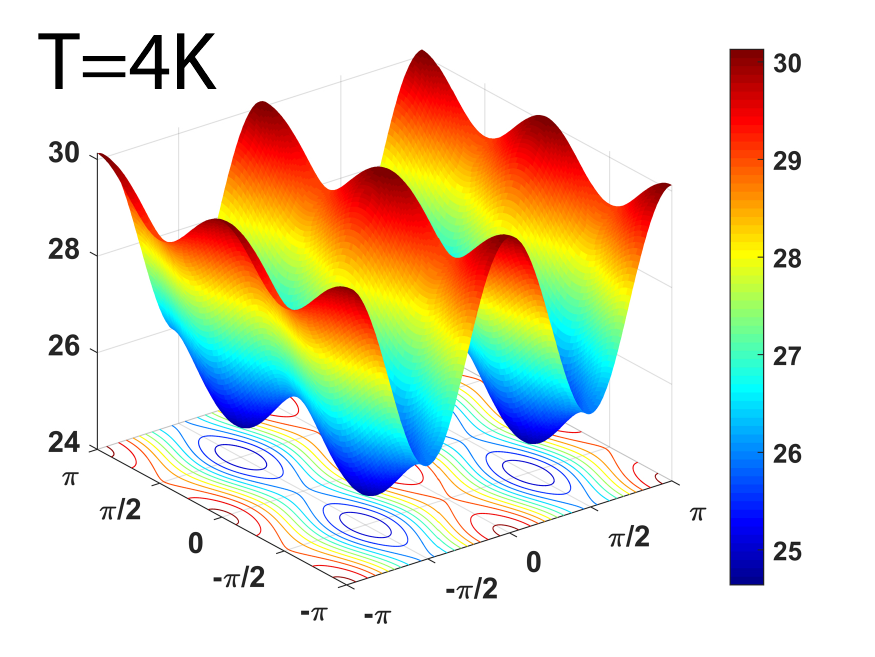}
    \caption{Above plots correspond to the dual single-mode squeezed input simulations (table \ref{t1}) with $r_1=r_2=0.8$.  Without squeezed input light, the quantum to thermal noise ratio greatly determined the cavity's entanglement output. When $r>0$, this somewhat holds, but the overall increase in entanglement is much less than when unsqueezed coherent light is input. In the two plots, the quantum to thermal noise ratio is maximum at the input squeezing angles where output entanglement is maximum (at about $(n\pi,m\pi)$ where $n$ and $m$ are some arbitrary integers); these findings are in full agreement with the entanglement simulation results.} 
    \label{fig:sqz_qnoise}
\end{figure}

While entanglement generation methods' advantages will vary with system parameters, the maximal overall benefit provided by OM methods to entanglement production only seems slight at first glance. This middling benefit to entanglement is seemingly in contrast to the significant increase in the quantum noise (see figure~\ref{fig:sqz_qnoise}). The expectation is that the quantum noise drives (or is an indicator of) the optomechanical entanglement. Figure~\ref{fig:sqz_qnoise} shows notable quantum variances that are at most two orders of magnitude higher than the unsqueezed input variances \cite{Kahlil2020}, yet the output entanglement seems only weakly boosted.

The output entanglement could be increasing proportionally to the increase in input quantum noise, but the increase in the number of entangled modes obscures the evidence in our measurements. Concurrently, the temperature dependence is greatly lessened. This effect was also found to be the case in other publications with squeezed light injection into OMCs \cite{AuBougouffa2020}. This is likely due to the squeezing changing the OMO's zero-point fluctuations or temperature, and this effect should be investigated further in the ponderomotive regime. Thus, we observe an increase in output entanglement is about six times the entanglement output instead of about 30 times the output (as is the case for the quantum to thermal noise ratio). 

\subsection{Squeezing Resources Usage Efficacy}


The simulations show OMC-based circuits have the potential for applications as entangling devices that rival conventional means of entanglement production. Consider situations where losses are low and nonlinear crystals of high nonlinearity and length, are scarce. 
Then in a setup with two lasers of set output powers $p_1$ and $p_2$, the maximum entanglement produced with the nonlinear crystal is proportional to the amount of squeezing the can be obtained from the crystal and light sources. The squeezing strength relates to the pump power and the effective non-linearity $\mu$ as $r = \mu \sqrt{p_1+p_2}$ \cite{Kaiser:16}. Our laboratory simulations take $p_1=46.24\mu $W and $p_2=1.1$mW (note that if $\mu_1=\mu_2$ then $r_1\neq r_2$ and if $r_1=r_2$ then $\mu_1\neq \mu_2$). Using these, we can plot (figure~\ref{fig:mutheta1}) the entanglement from conventional methods and compare it to entanglement from methods that inject squeezed light into the optomechanical cavities (crystal + optomechanical entanglement method); to better consider means of producing entanglement are already available and demonstrates the advantage of OMC integration. 
Duan's measure is sufficient but not necessary for non-Gaussian states. It is not reliable in comparing entanglement output between entangling devices if one or both produce non-Gaussian entanglement.


\FloatBarrier

\section{Conclusions}
Entanglement and squeezing are valuable quantum resources in quantum technologies. Quantum Cavity Optomechanics contains new means of producing squeezed and entangled light. Generating entanglement via squeezed light injection into sideband unresolved OMCs may well be a competitive method for producing, enhancing, or processing entanglement from single-mode squeezing resources for continuous-variable quantum devices. These simulations show a non-trivial relationship between input squeezing, intracavity entanglement, and intracavity quantum noise. At the right squeezing angles, implementing an optomechanical cavity in a conventional entanglement circuit could increase output entanglement, quantified by Duan's measure of separability, by a factor of 2 or greater due to the increase in the injected quantum noise. These results are consistent in our theoretical results (Quantum Langevin) and our simulations (two-photon formalism quadrature propagation). The increase in quantum noise due to squeezing helps mitigate unwanted effects due to thermal noises, as was the case in previous studies \cite{AuBougouffa2020}.
Lastly, the intracavity entanglement can be controlled by controlling the squeezing angle of the input light driving the system into the region of increased bipartite entanglement.

\section*{Acknowledgement}
KD, LC, and NB dedicate this paper to the memory of their former supervisor, late Professor Jonathan P. Dowling. KD, LC, NB, and JPD would like to acknowledge the Air Force Office of Scientific Research, the Army Research Office, the Defense Advanced Research Projects Agency, and the National Science Foundation.  We would also like to thank Professor Mark Wilde for important discussions. This material is based upon work supported by the National Science Foundation under Grant No. PHY-1806634.

\section{Appendix}
\FloatBarrier
\subsection{Computational methods}
Much of the computational framework from our previous project was preserved for this work. 
Here the matrix $\mathscr{S}$ describes how the sideband operators transform due to single-mode squeezing
\begin{eqnarray}
\mathscr{S}(r,\theta)= ~~~~~~~~~~~~~~~~~~~~~~~~~~~~~~~~~~~\\
\left(
\begin{array}{cc}
 \cos{2\theta} \sinh{r}+\cosh{r}
    & \sin{2\theta} \sinh{r} \\
 \sin{2\theta} \sinh{r} &
   \cosh{r}-\cos{2\theta}
   \sinh (r) \\
\end{array}
\right)\nonumber~. 
\end{eqnarray}
Since the program assumes squeezing of the sideband fields prior to injection, none of the squeezing is lost due to the cavity input coupling.

The two-photon formalism's flexibility allows for the thorough simulation of the sideband quadratures' homodyne measurement \cite{PhysRevA.72.013818,dixon_2021,Kahlil2020,Aggarwal2020,cripe_2018,sharifi2019design}. Such flexibility creates a robust platform for understanding how the output entanglement and quantum noise are affected by laboratory complications and loss parameters.

\subsection{Squeezing angle dependence}
In this section we note the steps to calculate the expected value of $\hat{\mathbf{S}}^\dagger(\xi)\hat{\mathbf{S}}(\zeta)$. Before we begin we note that $\hat{\mathbf{S}}^\dagger(\xi)\hat{\mathbf{S}}(\zeta) = \hat{\mathbf{1}}$ when $\xi=\zeta$ and $\hat{\mathbf{S}}^\dagger(\zeta)\hat{\mathbf{S}}(\xi)= \hat{\mathbf{S}}(\zeta)$ when $ \xi = 2\zeta$ (will be shown in the following notes). 
Let's begin with the following:
\begin{eqnarray}
\hat{\mathbf{S}}^\dagger(\xi)\hat{\mathbf{S}}(\zeta)= e^{-\xi^* \hat{\mathbf{a}}^2+\xi \hat{\mathbf{a}}^{\dagger2} } ~ e^{\zeta^* \hat{\mathbf{a}}^2 -\zeta \hat{\mathbf{a}}^{\dagger2} }
\end{eqnarray}
Using the Baker-Campbell-Hausdroff (BCH) formula we can write this product as follows
(note: $\mathbf{A}= -\xi^* \hat{\mathbf{a}}^2+\xi \hat{\mathbf{a}}^{\dagger2}$ and $\mathbf{B}=\zeta^* \hat{\mathbf{a}}^2 -\zeta \hat{\mathbf{a}}^{\dagger2}  $ ):

\begin{equation}
\begin{array}{r}
\hat{\mathbf{S}}^{\dagger}(\xi) \hat{\mathbf{S}}(\zeta)=e^{-\xi^{*} \hat{\mathbf{a}}^{2}+\xi \hat{\mathbf{a}}^{\dagger 2}} e^{\zeta^{*} \hat{\mathbf{a}}^{2}-\zeta \hat{\mathbf{a}}^{\dagger 2}} \\
=\exp \left[\left(-\xi^{*}+\zeta^{*}\right) \hat{\mathbf{a}}^{2}+(\xi-\zeta) \hat{\mathbf{a}}^{\dagger 2}+\hat{\mathbf{Z}}\right]
\end{array}
\label{eqnBCH0}
\end{equation}


where $\hat{\mathbf{Z}}= \frac{1}{2}\hat{\mathbf{k}}_0+ \frac{1}{12}( [\hat{\mathbf{f}}_2(-\xi),\hat{\mathbf{k}}_0]- \frac{1}{12} [\hat{\mathbf{f}}_2(\zeta),\hat{\mathbf{k}}_0])+...$ with $\hat{\mathbf{k}}_0= \frac{1}{2}[-\xi^* \hat{\mathbf{a}}^2+\xi \hat{\mathbf{a}}^{\dagger 2} ,\zeta^* \hat{\mathbf{a}}^2-\zeta \hat{\mathbf{a}}^{\dagger 2}]$.

\begin{equation}
\begin{array}{l}
\hat{\mathbf{k}}_{0}=\left[\xi \hat{\mathbf{a}}^{\dagger 2}-\xi^{*} \hat{\mathbf{a}}^{2}, \zeta^{*} \hat{\mathbf{a}}^{2}-\zeta \hat{\mathbf{a}}^{\dagger 2}\right] \\
=\left[\xi \hat{\mathbf{a}}^{\dagger 2}, \zeta^{*} \hat{\mathbf{a}}^{2}\right]+\left[\xi^{*} \hat{\mathbf{a}}^{2}, \zeta \hat{\mathbf{a}}^{\dagger 2}\right]
\end{array}
\label{eqnk0}
\end{equation}


since, $[\hat{\mathbf{a}}^{\dagger 2}, \hat{\mathbf{a}}^2]= -2 (2\hat{\mathbf{n}}+1)$ and $[\hat{\mathbf{a}}^2,\hat{\mathbf{a}}^{\dagger 2}]= 2 (2\hat{\mathbf{n}}+1)$ we can write eqn \ref{eqnk0} as
\begin{eqnarray}
2(\xi \zeta^*-\xi^*\zeta)(2\hat{\mathbf{n}}+1)~.
\end{eqnarray}

\begin{figure}[ht!]
    \centering
    \includegraphics[width=8.2cm]{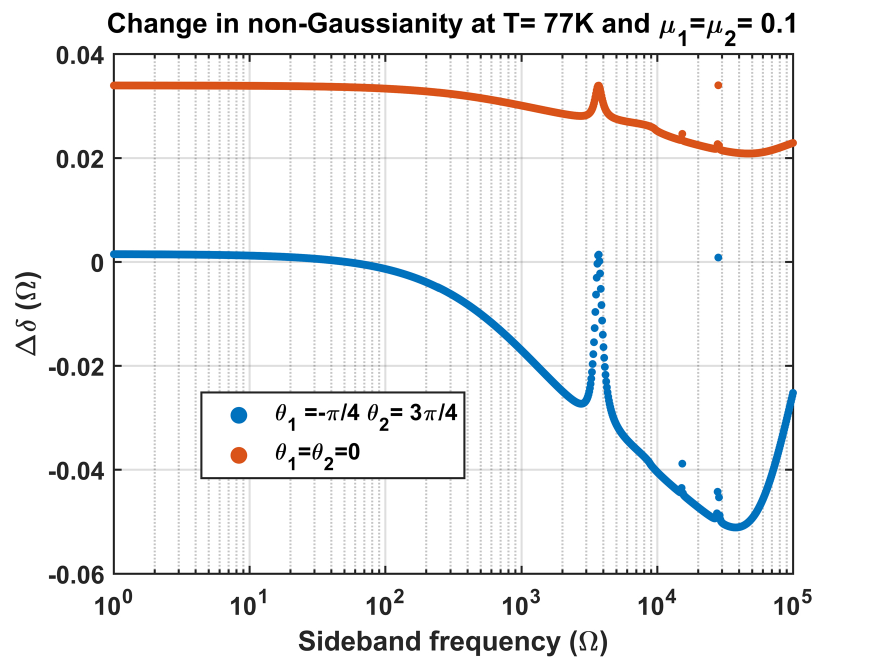}
    \includegraphics[width=4.1cm]{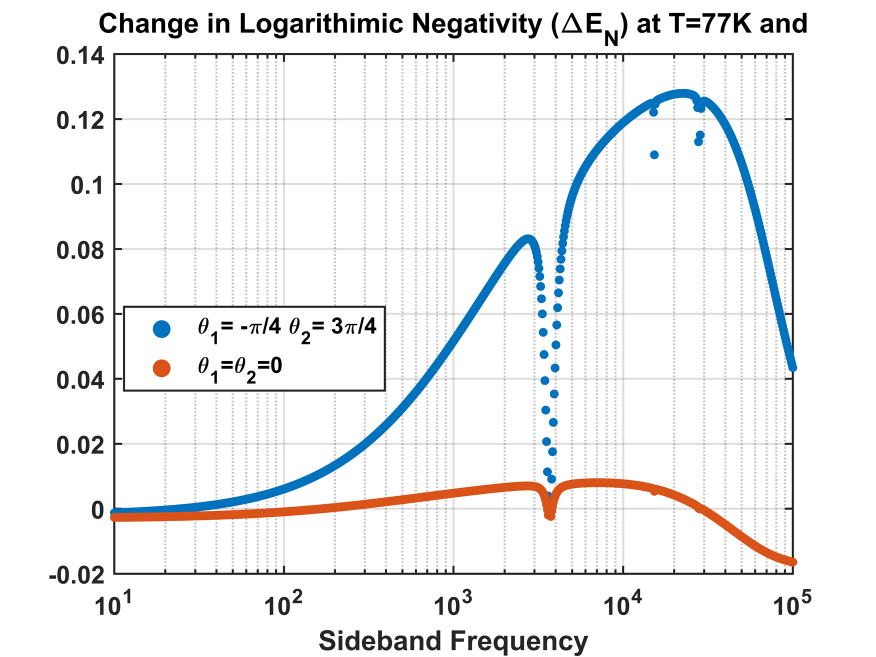}\includegraphics[width=4.1cm]{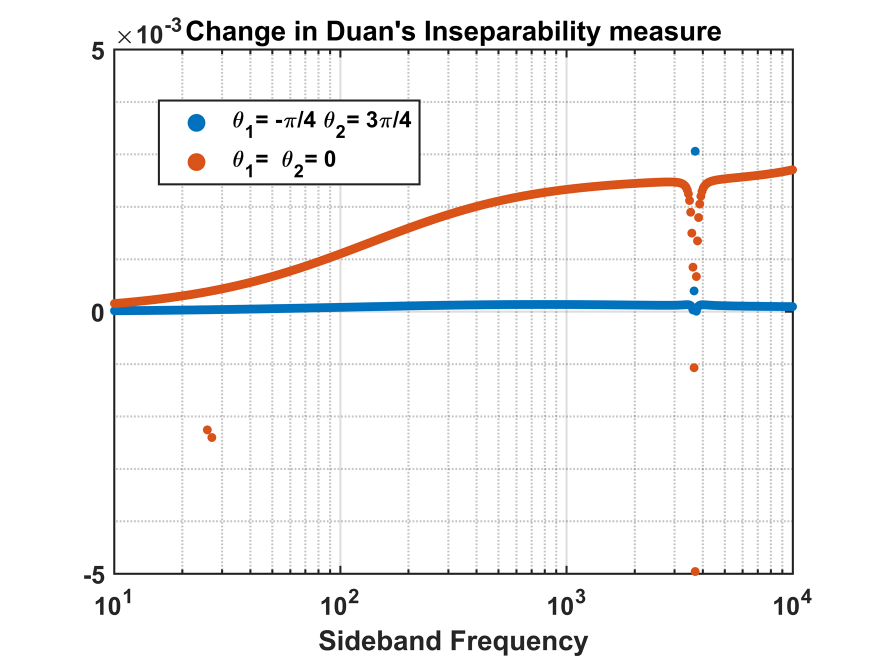}
    \caption{These plots examine changes in Gaussianity measure and entanglement changes with small squeezing ($\mu=0.1$ relative to unsqueezed input) versus sideband frequency, $\Omega$, at 77K. }
    \label{fig:freq}
\end{figure}

Thus, the resulting operator is highly dependent on the input squeezing angle difference and is not very sensitive to differences in squeezing strength between the two inputs. To make this more clear we can write $\mathbf{\hat{k}}_0$ in a different form than above if we assume $\xi= r_1 e^{i\theta}$ and $\zeta= r_2 e^{i\phi}$:
\begin{eqnarray}
\hat{\mathbf{k}}_0~=~4i r_1r_2\sin{(\theta-\phi)}(2\hat{\mathbf{n}}+1)~.
\end{eqnarray} If the squeezing angles are the same, the term $\hat{\mathbf{k}}_0$ vanishes; the real parts of the squeezing parameters do not change the form of the operator unless they are zero. Furthermore, at higher orders there are terms with dependence on $\sin(\theta+\phi)$ and $\cos(\theta+\phi)$; there is still some non-trivial complications to the squeezing operator $around$ the line $\theta=\phi$. These complications may account for the peaks in the dual single-mode squeezed input entanglement results which appear to show that the entanglement is maximum when $\theta \approx \phi$; it is likely that when $\theta$ exactly equals $\phi$ that the $E_N$ falls (or rises) sharply.

\subsection{\label{sec:freq}Frequency dependence}


Considering the results in figure \ref{fig:Gaussianity} with those depicted in figure \ref{fig:freq}, the input squeezing, at the angles that yield less information loss, does not change the separability in Duan's measure; meaning there is no net change in output entanglement at these squeezing angles. In this configuration, the input squeezing manipulates tripartite entanglement to bipartite entanglement.

\begin{figure}[ht!]
    \centering
    \includegraphics[width=8.2cm]{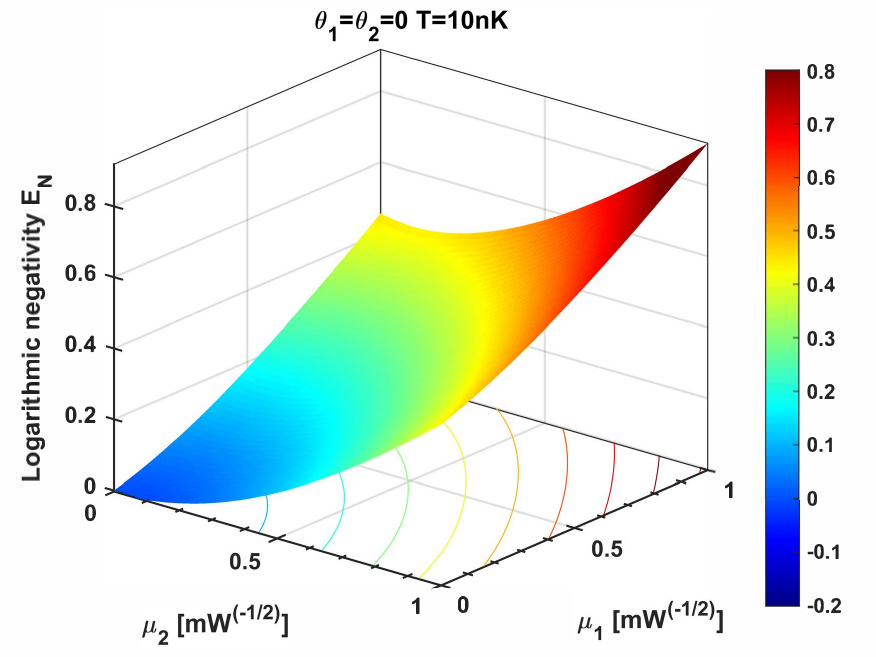}
    \includegraphics[width=4.1cm]{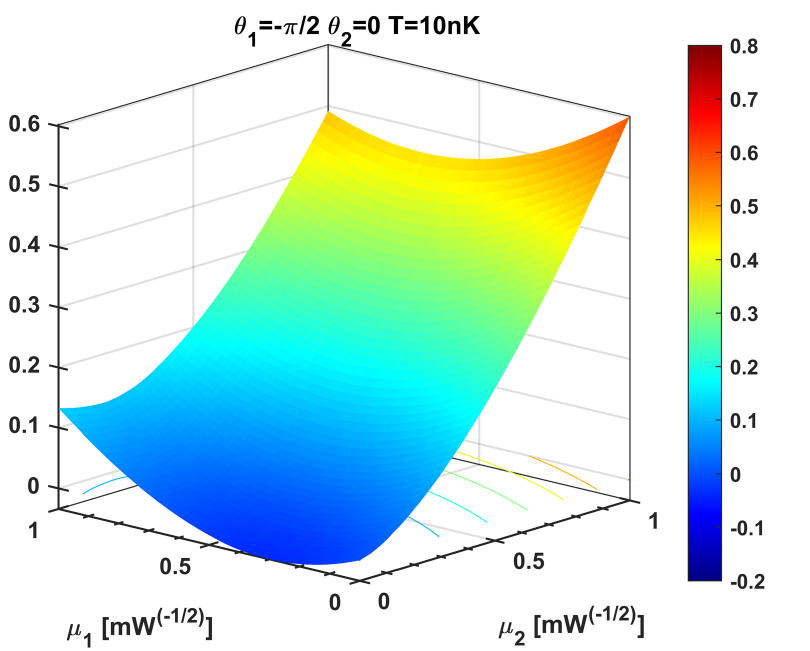}\includegraphics[width=4.1cm]{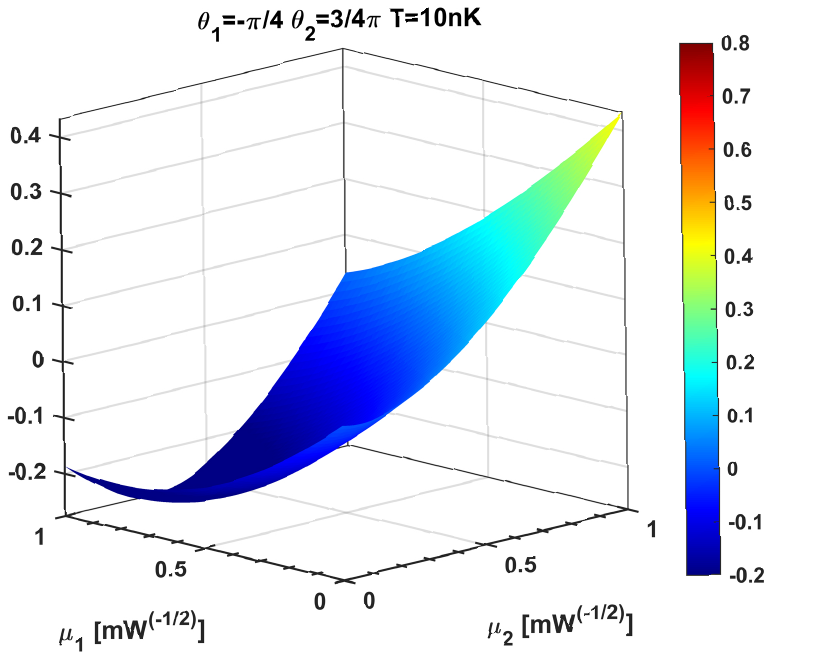}
    \caption{The above plots are of the changes in the Gaussianity metric versus squeezing strength. These ultimately show that increasing the entanglement with the OMO via input squeezing works best for low squeezing strengths (here $\mu >0.5$) or unbalanced squeezing.}
    \label{fig:Gauss_vs_mu}
\end{figure}

\subsection{Separating the tripartite state with input squeezing}
\label{sec:AppenGauss}
In our simulations, intense squeezing consistently detangles the optical fields from the OMO, see figure \ref{fig:Gauss_vs_mu}.

\FloatBarrier
\bibliography{mybib2}
\end{document}